\documentclass[epsfig,12pt]{article}
\usepackage{epsfig}
\usepackage{graphicx}
\usepackage{array}
\usepackage{color}
\usepackage{bm}
\usepackage{amsmath,amssymb,calc}
\usepackage{braket}
\usepackage{slashed}
\usepackage{bbold}
\usepackage{xcolor}

\newcommand{\beq}{\begin{equation}}   
\newcommand{\eeq}{\end{equation}}
\newcommand{\beqn}{\begin{eqnarray}}   
\newcommand{\eeqn}{\end{eqnarray}}

\def\ntwo{${\mathcal N}=2\;$}

\def\none{${\mathcal N}=1\;$}

\newcommand{\de}{{\rm Det}}
\newcommand{\pt}{\partial}

\newcommand{\lcal}{{\mathcal L}}

\newcommand{\jcal}{{\mathcal J}}
\newcommand{\qcal}{{\mathcal Q}}
\newcommand{\gsim}{\lower.7ex\hbox{$
\;\stackrel{\textstyle>}{\sim}\;$}}
\newcommand{\lsim}{\lower.7ex\hbox{$
\;\stackrel{\textstyle<}{\sim}\;$}}
\usepackage{showlabels}

\begin{document}

\begin{flushright}
FTPI-MINN-21-05, UMN-TH-4012/21
\end{flushright}

\begin{center}
{  \large \bf Bringing Yang-Mills  Theory 
 Closer to  Quasiclassics}
 
\end{center}

\begin{center}
{\Large Mikhail Shifman} 

\vspace{3mm}

{\it  William I. Fine Theoretical Physics Institute,
University of Minnesota,
Minneapolis, MN 55455, USA}
\end{center}

\vspace{2mm}

\begin{center}
{\large\bf Abstract}
\end{center}

{ A deformation of pure Yang-Mills theory by a phantom field similar to the Faddeev-Popov ghost is considered.
In this theory an {\em Ersatz}-supersymmetry is identified which results in cancellation of quantum corrections up to two-loop order.
A quadruplet built from two complex fields in the adjoint -- the Faddeev-Popov ghost $c^a$ and the phantom $\Phi^a$, all with the wrong statistics --
balances four gauge fields $a_\mu^a$.
At this level, the instanton measure and the 
$\beta$ function is fully determined by quasiclassics. In a simple $\phi^4$ theory with a phantom added I identify a strictly conserved  {\em Ersatz}-supercurrent.
In the latter theory unitarity of amplitudes persists despite the presence of the phantom. In deformed Yang-Mills it is likely (although not proven) to persist too in all amplitudes with only gluon external legs. It remains to be seen whether this construction is just a  device facilitating some loop calculations or broader applications can be found.}

\section{Introduction}

In the recent publication \cite{Shifman2020} I established a certain proximity between the $\beta$-function calculations in supersymmetric and non-supersymmetric versions of the Yang-Mills theory. This was achieved by changing statistics for certain fields, for instance, the ``second" gluino and its scalar \none superpartner in  \ntwo SYM were converted into the so-called {\em phantoms}. It was a purely technical computational device.

 The present paper,  although involving some of the ideas of the previous one, pursues a different goal. The question addressed is as follows:  
 Can we minimally deform {\em non}supersymmetric Yang-Mills to bring it closer to {\em exact}  results inherent to supersymmetric Yang-Mills theory (at least, in some aspects) without introducing spin-$\frac 12$ fermions?
 
 As we will see below, the answer to this question is positive. To explain how this happens let us start from  pure Yang-Mills theory 
 and add one {\em complex} scalar field $\Phi^a$  in the adjoint representation (for definiteness I will limit myself to the SU$(N)$ gauge group),
\beq
{\mathcal L}_{\rm YM\, adj\, ph}=-\frac{1}{4} G_{\mu\nu}^a  G^{\mu\nu\,a} + |D_\mu \Phi^a|^2
\label{P1}
\eeq
where $D_\mu$ is the standard covariant derivative.
Next, I change the statistics of  $\Phi^a$ so that in the path integral it will produce determinant in the numerator rather than in the denominator, \'a la the ghost field routinely introduced in Yang-Mills for the purpose of gauge fixing. Then in the free field theory limit 
($g=0$)   we have
\beq
{\mathcal L}_{\rm YM\, adj\, ph}=-\frac{1}{4} \left(\pt_\mu A^a_\nu - \pt_\nu A^a_\mu\right)^2 + \pt_\mu\overline{\Phi^a} \,\pt^\mu {\Phi^a}
\eeq
and, after the standard gauge fixing, the  Lagrangian above takes the form 
\beq
{\cal L} =  \bigg\{ -\frac 12 \left(\pt_\mu a_\nu^a\right)^2
+\left( \pt_\mu \bar c\right)\left( \pt^\mu c  \right) +\left(\pt_\mu \bar \Phi \right)\left(  \pt^\mu \Phi \right)
 \bigg\}\,.
 \label{P3}
\eeq 
It is obvious that the Lagrangian (\ref{P3}) is supersymmetric in this formulation: four bosonic degrees of freedom $a_\mu$ are balanced by four fermionic, $c,\,\bar c,\, \Phi,\,\bar\Phi$.
In particular, with the Lagrangian (\ref{P3}) the vacuum energy density automatically vanishes.

This does not take us too far, however, since free field theories by themselves are not interesting.
The question is whether this {\em Ersatz}-supersymmetry can be preserved, at least to some degree,  when we switch on gauge interactions i.e. consider $g\neq 0$.
Although I do not have the full answer, I will show that in one and two loops supersymmetry of (\ref{P3}) is maintained. 
Calculation of the instanton measure becomes fully quasiclassical  --  quantum corrections cancel in the first and second loops which allows one to  derive an analogue of the  NSVZ \none $\beta$ function. The presence of the phantom field in the theory (\ref{P1}) most likely does not spoil unitarity of pure Yang-Mills theory in amplitudes 
with the gluon external legs.

That this is indeed the case is demonstrated in a $\phi^4$ theory (see Section~\ref{phifour}) deformed by one additional phantom field. In this model it is easy to write an exactly conserved {\em Ersatz}-supercurrent. 

Whether or not the suggested construction is just a device for loop calculations, or more  applications can be found, remains to be seen.

In the past, somewhat similar spin-0 phantom fields appeared in \cite{PS} in the context of stochastic quantization.

\section{Background field}

 To separate the external field from the quantum gauge fields propagating in  loops we can always write
 \beq
 A_\mu^a \equiv \left(A_\mu^a\right)_{\rm ext} +a_\mu^a\,,
 \label{c119}
 \eeq
  Substituting (\ref{c119}) in the gluon part of the Lagrangian (\ref{P1}) and keeping only the terms {\em quadratic} in $a_\mu^a $ we obtain
 \beqn
 {\cal L} &=& \frac{1}{g^2}\left\{-\frac 14 \left( G_{\mu\nu}^a G^{\mu\nu\, a}\right)_{\rm ext} -\frac 12 \left(D_\mu^{\rm ext} a_\nu^a\right)^2
 +\frac 12  \left(D_\mu^{\rm ext} a^{a\,\nu}\right) \left(D_\nu^{\rm ext} a^{a\,\mu}\right)\right.
  \nonumber\\[2mm]
 &+& \frac 12 a^{a\,\mu}  \left(G_{\mu\nu}^b\right)_{\rm ext}f^{abc} a^{c\,\nu}
 \Big\}+ ...
 \label{c1114p}
\eeqn
 The action of  $D_\mu^{\rm ext} $ on $a_\mu$ is defined as
 \beq
 D_\mu^{\rm ext} a_\nu^a (x) \equiv \partial_\mu\, a_\nu^a -i  \left(A_\mu^b\right)_{\rm ext} \left(T^b\right)^{ac} a_\nu^c
 \equiv \partial_\mu\, a_\nu^a +f^{abc} \left(A_\mu^b\right)_{\rm ext} a_\nu^c\,.
 \label{20p}
 \eeq  
 Next, we impose the gauge condition
  \beq
 D_\mu^{\rm ext} a^{a\, \mu} = 0\,,
 \eeq
 rather than  the standard $\partial^\mu A_\mu^a = 0$. The Faddeev-Popov ghost determinant can be written in terms of the ghost fields 
  $\bar{c},\, c$
as follows:
 \beq
  {\cal L}_{\rm ghost} =  \left( D_\mu^{\rm ext} \bar c\right)\left(  D^{\mu\,\rm ext}  c  \right)\,.
  \label{c1117}
 \eeq
 The ghost fields above are scalar complex fields in the adjoint representation of the gauge group and ``wrong'' statistics. 
 Note that in this formalism the derivatives acting on $c$ and $\bar c$ enter symmetrically.
 
 After adding the gauge fixing term
 \beq
\Delta  {\cal L}_{\rm gauge} = -\frac{1}{2g^2} \left(D_\mu^{\rm ext} a^{a\,\mu} 
\right)^2\,.
\label{c1118}
\eeq
and integrating by parts
 the quantum part of the Lagrangian (\ref{P1}), (\ref{c1114p}), and (\ref{c1117}) takes the form
  \beqn
 {\cal L}_{\rm quant} & =&  \frac{1}{g^2}\bigg\{ -\frac 12 \left(D_\mu^{\rm ext} a_\nu^a\right)^2
+a^{a\,\mu}  \left(G_{\mu\nu}^b\right)_{\rm ext}f^{abc} a^{c\,\nu} \nonumber\\[2mm]
&+&\left( D_\mu^{\rm ext} \bar c\right)\left(  D^{\mu\,\rm ext}  c  \right) +\left( D_\mu^{\rm ext} \bar \Phi \right)\left(  D^{\mu\,\rm ext}  \Phi \right)
 \bigg\}\,.
 \label{c1120}
\eeqn
The BRST symmetry guaranties that it is equivalent to (\ref{P1}). An SU(2) symmetry with regards to $c\Phi$ rotations is obvious in the second line.

\begin{figure}[h]
\centerline{\includegraphics[width=5cm]{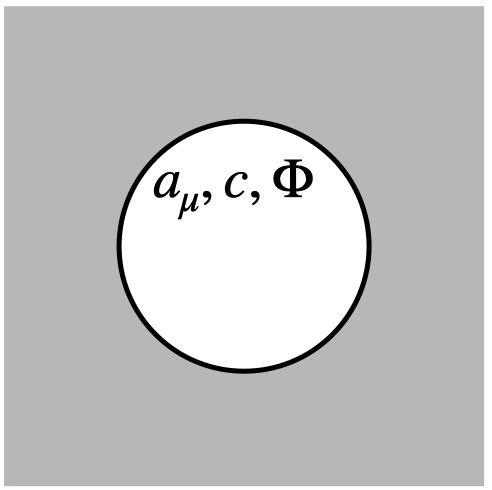}}
\caption{\small The sum of three loops vanishes provided that we ignore the gluon spin interaction (i.e. the second term in
(\ref{c1120})). The gray area denotes the background field in which $a_\mu$, $c$, and $\Phi$ propagate.}
\label{alphag} 
\end{figure}

The phantom field $\Phi$ acts exactly as the second ghost.
 We immediately see that at one loop the first term in (\ref{c1120}) is completely cancelled by two terms in the second line.
 It is only the spin interaction of the gluon fields (the second term in the first line) that survives, see Fig. \ref{alphag}.

\section{Instanton background}

As well-known from the pioneering 't Hooft's paper \cite{hooft} (see his discussion after Eq. (4.6)), all three operators appearing in (\ref{c1120}) have coinciding
eigenvalues with the only exception -- the zero modes which are present for the vector fields (due to their magnetic interactions) but
are absent for the scalar fields $\Phi$ and $c$\,. Therefore, the appropriate product of three determinants (gauge, ghost and the phantom fields $\Phi$)
reduces to unity,
\beqn
&&\left\{\de^\prime\left[ -\frac 12 \left(D_\mu^{\rm ext} \right)^2+  \left(G_{\mu\nu}^b\right)_{\rm ext}\varepsilon^{abc}   \right]_a\right\}^{-1}
\nonumber\\[2mm]
&\times& \de\left[ - \left(D_\mu^{\rm ext} \right)^2_c\right] \times  \de\left[ - \left(D_\mu^{\rm ext} \right)^2_\Phi\right] = 1\,,
\label{P12}
\eeqn
provided the number of degrees of freedom matches -- which it does, see Eq. (\ref{P3}). The prime in the first determinant means that the zero modes are removed. There are no zero modes in the other two determinants.

The instanton measure in  SU$(N)$ Yang-Mills theory  is 
\beq
d\mu_{\rm inst}  ={\rm const}\times 
\, \int \frac{d^4x_0\, d\rho}{\rho^5} \big(M_{\rm uv}\rho \big)^{4N} \left( \frac{8\pi^2}{g^2}\right)^{2N}
\, \exp\left(- \frac{8\pi^2}{g^2}  + \Delta_{\rm gl} + \Delta_{\rm gh}
\right),
 \label{4553}
\eeq
where the exponent ${8\pi^2}/{g^2}$ is the action of  the classical solution, other pre-exponential factors are due to zero modes,
while $\Delta_{\rm gl}+ \Delta_{\rm gh} $ describes one-loop {\em bona fide} {\sl quantum} correction due to gluons and ghosts in the instanton background,
\beq
\Delta_{\rm gl}+ \Delta_{\rm gh} = -\frac 13 N \log \big(M_{\rm uv}\rho \big)\,.
\label{ququ13}
\eeq
The zero modes emerge due to the spin term (i.e. the second term in (\ref{c1120})). 

Let us add now the contribution of the fourth term corresponding to our newly added phantom field.
Integrating out $\Phi$ we obtain
\beq
{\rm Det} \left( - D_\mu^{\rm ext} D^{\mu\,\rm ext}   \right) \to \exp \left\{{\rm Tr }\log  \left( - D_\mu^{\rm ext} D^{\mu\,\rm ext}   \right)\right\}\,.
\label{P15}
\eeq
Because of the phantom nature of $\Phi$ its contribution is presented by Det rather than Det$^{-1}$.
The answer is of course known from the studies of a regular adjoint scalar field, we have just to change the overall sign. The result
reduces to $e^{\Delta_\Phi}\,,$
\beq
 \Delta_\Phi = \frac 13 N \log \big(M_{\rm uv}\rho \big) = - \left( \Delta_{\rm gl}+\Delta_{\rm gh}\right)\,. 
 \label{P13}
\eeq
Thus, $ \Delta_\Phi$ cancels $\Delta_{\rm gl}+ \Delta_{\rm gh} $ in Eq. (\ref{c1120}), see Eq. (\ref{P12}). The instanton measure is fully determined by 
the gluon zero modes.

In fact, there is no need to calculate  $\Delta$s at all.  As I have already mentioned, at one loop there is a supersymmetry which connects the 
$a_\mu$ contribution on one hand with that of $c$ and $\Phi$ combined, on the other hand.
Indeed, let us introduce a four-component phantom column
\beq
X^a_\rho =\left(\begin{array}{c}
{\rm Re}\, \Phi^a\\[1mm]
{\rm Im}\, \Phi^a\\[1mm]
{\rm Im} \, \,c^a\\[1mm]
{\rm Im}\, c^a\\[1mm]
\end{array}\right)
\eeq
Then the fantom quadruplet $X^a_\rho$ and four components of the vector fields $ a^a_\mu$  form a ``supermultiplet"  (in Euclidean formalism).
Of course we mix here two symmetries -- Euclidean Lorentz rotations and O(4) symmetry of $X_\rho$ rotations, but this has no impact on our final result.
The second line in (\ref{c1120}) is replaced by
\beq
\left( D_\mu^{\rm ext} \bar c\right)\left(  D^{\mu\,\rm ext}  c  \right) +\left( D_\mu^{\rm ext} \bar \Phi \right)\left(  D^{\mu\,\rm ext}  \Phi \right)\to
\frac{1}{2} \left( D_\mu^{\rm ext} X\right)\Gamma \left(  D^{\mu\,\rm ext}  X  \right)
\eeq
where $\Gamma$ is an appropriately chosen $4\times 4$ matrix $(\Gamma^2=1)$ of the type diag$\{\sigma_2, \sigma_2\}$. The fact that
$$
\de^\prime\left[  \left(D_\mu^{\rm ext} \right)_a^2\right]^{-1}\left(D_\mu^{\rm ext} \right)_X^2 =1
$$
is obvious. Below I will argue that this cancellation extends to two loops.
Then we can immediately  determine 
the two-loop beta function from a {\em purely classical calculation in a purely bosonic theory}.
Indeed, the instanton measure is
\beq
\big(M_{\rm uv}\big)^{4N} \left(\frac 1 {g^2}\right)^{2N}\exp\left( -\frac{8\pi^2}{g^2}\right) = {\rm RGI}
\label{P14}
\eeq
where the bare coupling constant in the exponent and in the pre-exponent has to conspire with $M_{\rm uv}$ to ensure that the left-hand side is
renormalization-group invariant.

\section{The \boldmath{$\beta$} function}

As a results  we arrive at
\beq
 \frac{d}{d \log M_{\rm uv}} \Big(4N\log M_{\rm uv} -2N\log g^2 - \frac{8\pi^2}{g^2} \Big)=0\,.
 \label{P19}
\eeq
If we define $\beta$ function as
\beq
\beta (\alpha) = \frac{d}{d \log M_{\rm uv}} \alpha =  -\beta_1\, \frac{\alpha^2}{2\pi} - \beta_2\, \frac{\alpha^3}{4\pi^2}+ ...\,,
\label{betaf}
\eeq
then we have
\beq
\beta_1=4N\,,\quad \beta_2 = 8N^2\,.
\label{P17}
\eeq
The above formula coincides with the standard perturbative calculation, provided we change the sign of the scalar adjoint loops to the one appropriate to
the scalar phantoms,
see e.g. \cite{jones} or  Eq. (A1) in Appendix of \cite{Shifman2020}. The structure of Eqs. (\ref{P14}) and (\ref{P19}) is the same as in the 
 NSVZ derivation  \cite{NSVZinst}.
 The presented derivation, being classical, is remarkably simpler than the standard perturbative two-loop calculation.
 
\begin{figure}[t]
\centerline{\includegraphics[width=7cm]{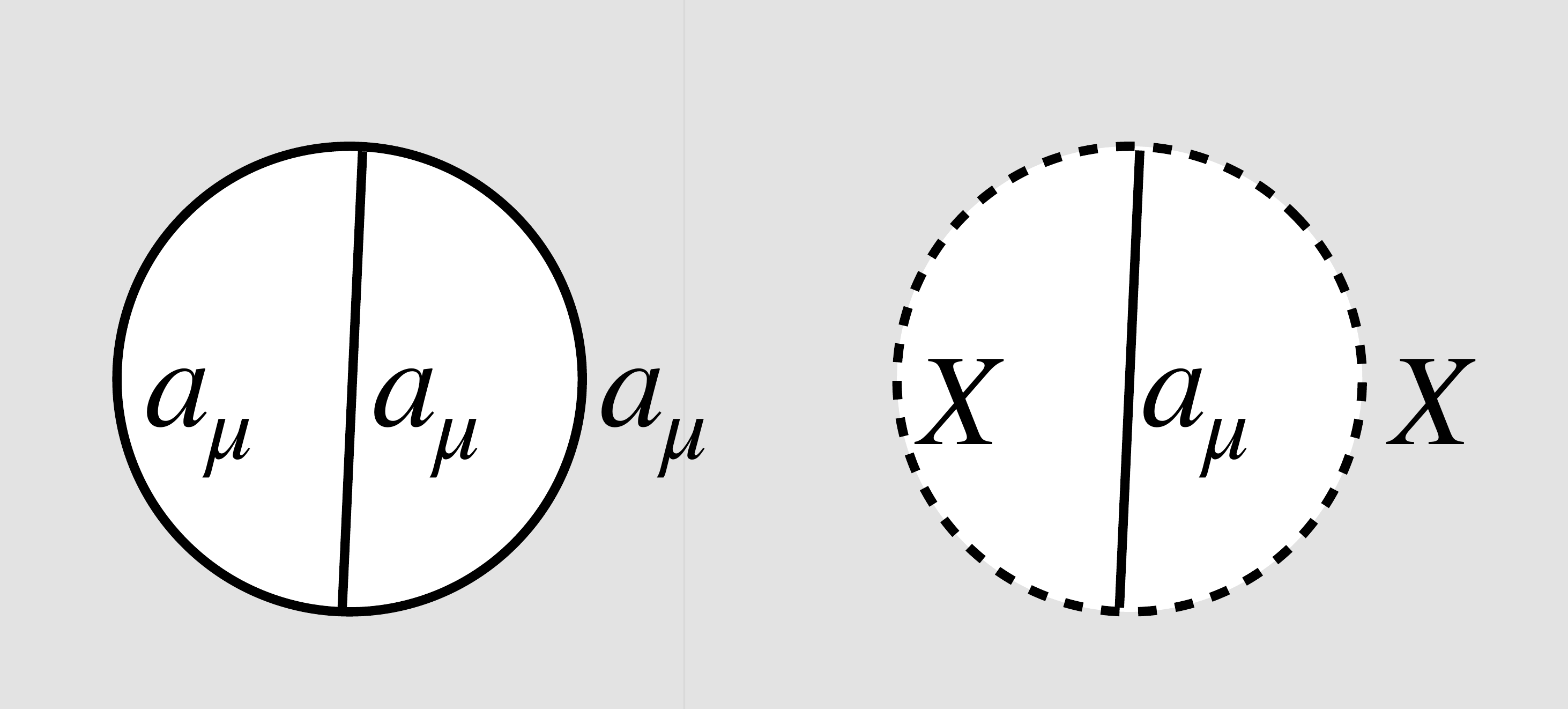}}
\caption{\small Two diagrams in the instanton background contributing at two loop-order.
They cancel each other due to the phantom nature of the $X$ field.}
\label{alp} 
\end{figure}

 Now I will argue that the second loop does not modify Eq. (\ref{P14}). We  add two graphs in the instanton background field, see Fig. \ref{alp}. They cancel each other because of $aX$ supersymmetry. It is probable that it can be extended to higher orders.
 
\section{How such theories can exist and be meaningful}
\label{phifour}

\begin{figure}[t]
\centerline{\includegraphics[width=7cm]{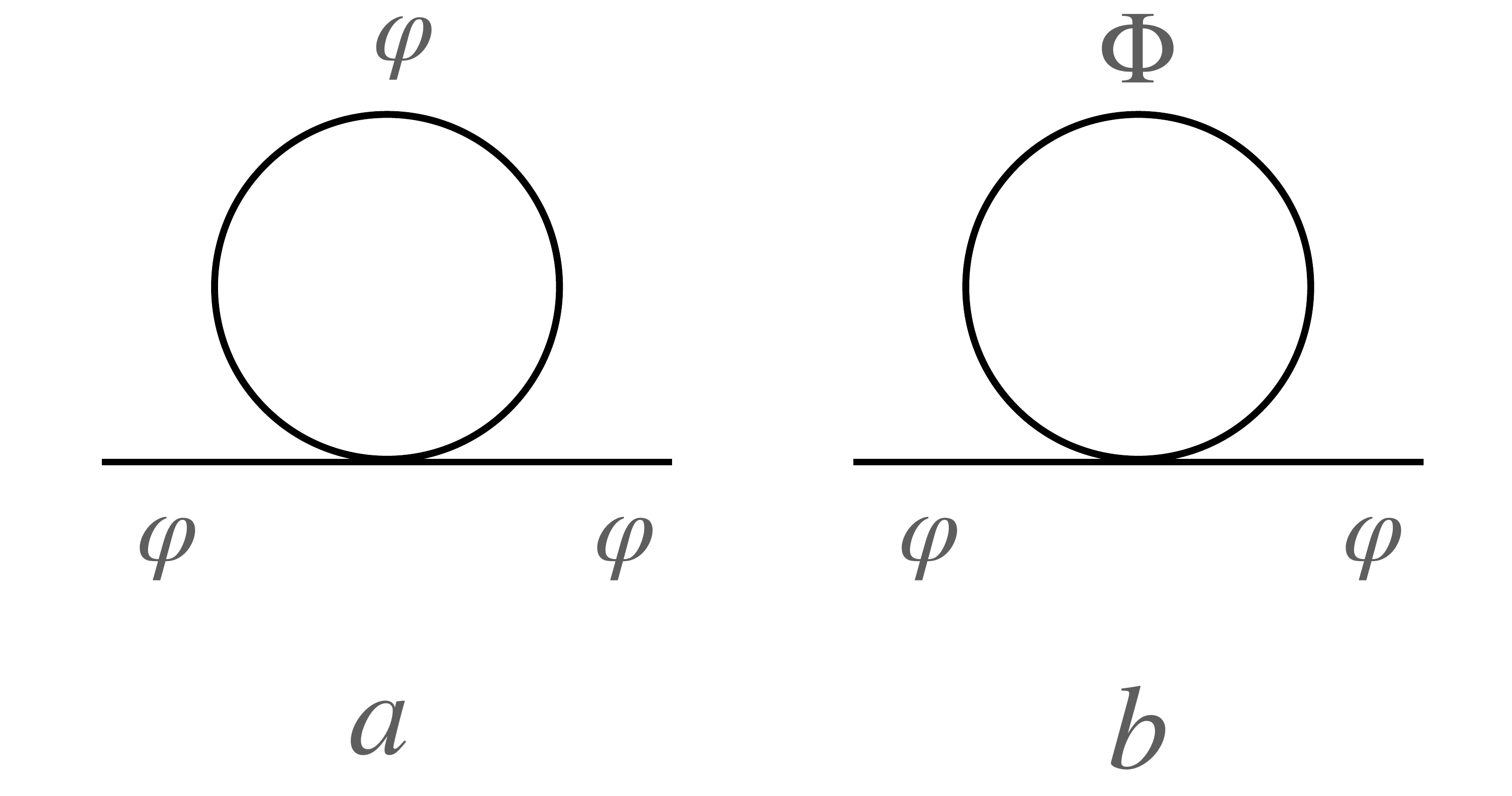}}
\caption{\small Tadpole contributions to the $Z$ factor of the $\varphi$ field.}
\label{tad1} 
\end{figure}

Here I would like to discuss more generic ``special supersymmetry" theories.
The simplest example I can think of is a scalar theory of the following pairs of complex fields:
$\varphi, \bar\varphi$ (regular fields) and $\Phi, \bar\Phi$ (phantom fields). The Lagrangian is
\beq
\lcal_{\varphi\Phi}= \pt_\mu\bar \varphi \pt^\mu\varphi  + (\bar\varphi\varphi)^2 + \pt_\mu\bar \Phi \pt^\mu\Phi  + 2 (\bar\varphi\varphi)  (\bar\Phi\Phi) \,.
\label{P22}
\eeq
The corresponding equations of motion take the form
\beqn
\pt^2\varphi &=& 2\varphi^2\bar\varphi  + 2\varphi (\bar\Phi\Phi),
\nonumber\\[2mm]
\pt^2\Phi &=&  2 (\bar\varphi\varphi) \, \Phi\,.
\eeqn
Then one can readily derive a conserved supercurrent,
\beqn
\jcal^\mu &=& \left(\varphi\stackrel{\leftrightarrow}{\pt^\mu} \Phi + {\rm H.c.}
\right),
\nonumber\\[2mm]
\pt_\mu\jcal^\mu &=&  \varphi (\pt^2\Phi)  - (\pt^2\varphi)\Phi =  2\varphi^2\bar\varphi\,\Phi   - 2\varphi^2\bar\varphi\,\Phi + \rm H.c.= 0\,.
\eeqn
Of course, the anticommutator $\{\qcal\jcal^\mu\}$ is unrelated to Hamiltonian; rather it is expressed through a combination of the U(1) bosonic currents,
\beq
\{\qcal\jcal^\mu\} =-i  \sum_{\phi= \varphi, \,\Phi} \, \phi \stackrel{\leftrightarrow}{\pt^\mu}\bar\phi\,.
\eeq
The theory (\ref{P22}) is not empty. Let us consider, for example, the simplest tadpole graphs depicted in Fig. \ref{tad1}.
Although the phantom loop in Fig. \ref{tad1}b cancels a part of the $\varphi$ loop in Fig. \ref{tad1}a this cancellation is not complete, with the result $4-2=2$ in appropriate units. The first number comes from the diagram Fig. \ref{tad1}a, the second from \ref{tad1}b. A similar tadpole 
for the $Z$ factor of the $\Phi$ field is represented by one graph which yields 2 in the same units.

\begin{figure}[t]
\centerline{\includegraphics[width=5.6cm]{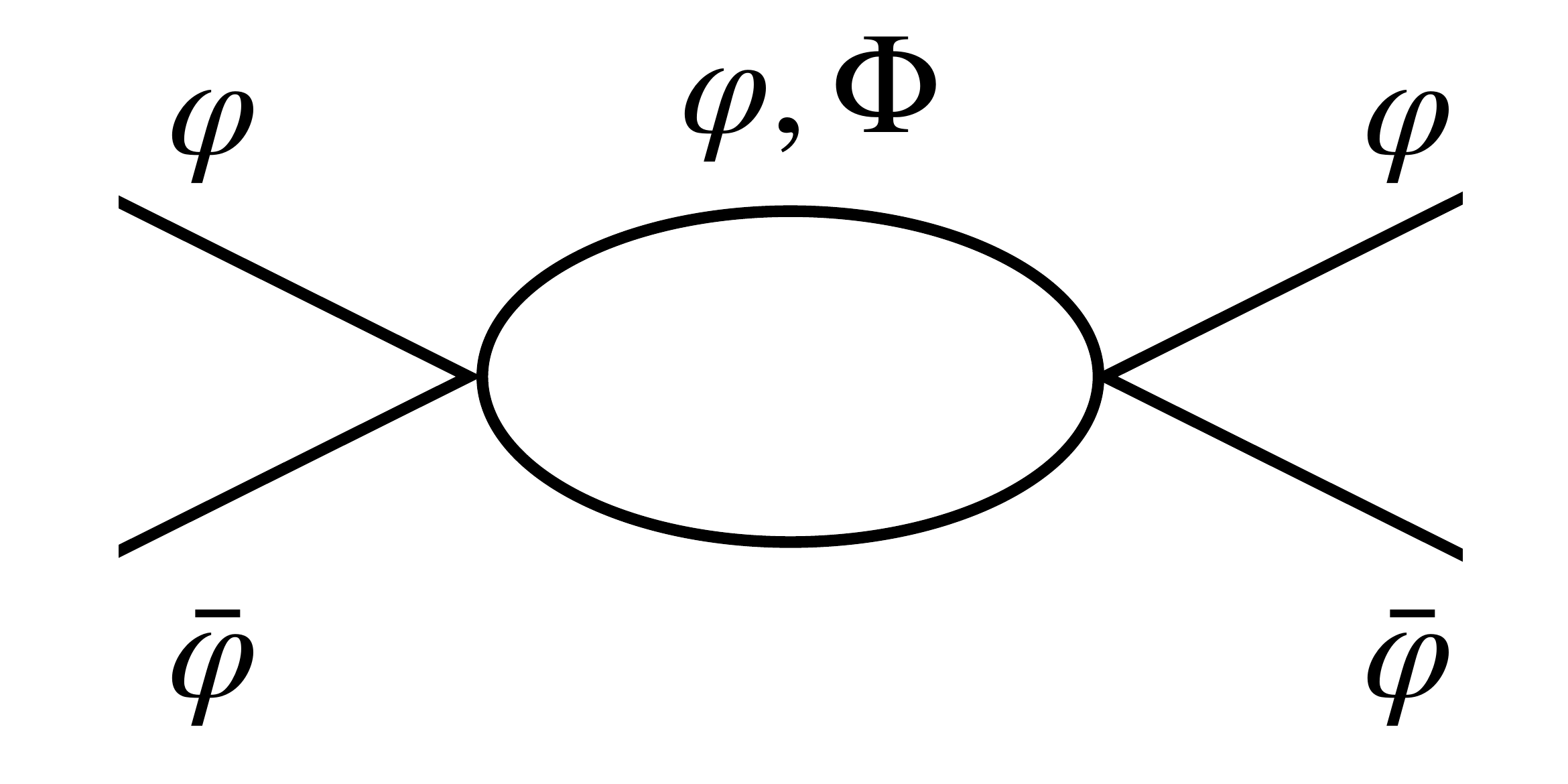}}
\caption{\small Two-by-two scattering amplitude at one  loop.
}
\label{four} 
\end{figure}

By the same token the negative $\Phi $ contribution to the two-by-two scattering graph in Fig. \ref{four} cancels just a small part of 
the positive $\varphi$ loop. The theory remains unitary.

\section{Conclusions}

If we add a complex scalar field $\Phi^a$  in the adjoint representation in pure Yang-Mills theory and then reverse its statistics then, in fact, we introduce
a certain supersymmetry in the theory thus obtained. The Faddeev-Popov ghost field combined with $\Phi^a$ creates a quadruplet acting as a superpartner for the vector field  $a_\mu$. The BRST symmetry is enhanced in this model. 

Limiting the physical sector to amplitudes in which $\Phi^a$ propagate only in loops
we get a theory which is probably unitary, with a simpler structure of the gluon interactions. For the time being the unitarity  statement must be viewed as a {\em conjecture}. 
One can give an argument in its favor, however.  Each gluon loop in Feynman graphs is accompanied by the same one with either the Faddeev-Popov ghosts or phantoms. If we disregard phantoms the remaining theory is just Yang-Mills theory which is certainly {\em unitary}.  Nonunitary contributions are due to the phantom
loops. Phantom interactions with gluons are of the charge type, just as those of the Faddeev-Popov ghosts. Let us generically refer to such interaction as $g_C$ while the magnetic interaction as $g_M$.

 As is seen from our previous analysis of the simplest graphs the contributions of the charge vertices are numerically suppressed compared to the vertices of the magnetic type roughly by the factor
  $$\frac{g_C^2}{g_M^2} \sim \frac 1 d$$
 where $d$ is the number of space-time dimenstons.
  The added phantoms cancel those 
gluon loop contributions  which have exclusively charge vertices. The contributions due to magnetic vertices (their minimum number is two) remain intact.  If this suppression persists in arbitrary graphs (still an open question) and if 
$d=4$ can be viewed as a large number, the theory with extra phantom field $\Phi^a$ will come out  unitary.

The instanton measure is determined at the (quasi)classical level and is similar to the NSVZ result (at least, up to two loops). Whether or not this construction can be extended to higher loops remains to be seen. If the answer is yes, then the exact result for $\Lambda^4$ in this theory would be
\beq
\Lambda^4 = M_{\rm uv}^4 (g^2)^{-2} \exp \left( -\frac{S_{\rm inst}}{N}\right)
\eeq
and the exact NSVZ $\beta$ function
\beq
\beta(\alpha )=-4N \, \frac{\alpha^2}{2\pi}\,\frac{1}{1- \frac{N\alpha}{\pi}}
\eeq
 will follow.
 
 \section*{Acknowledgments}
I am grateful to A. Cherman, G.  Dunne, and M. \"Unsal for useful communications.

This work is supported in part by DOE grant de-sc0011842.

%

\end{document}